\documentclass[twocolumn,preprintnumbers,amsmath,amssymb]{revtex4}
\usepackage{amsfonts}
\usepackage{graphicx}

\begin{document}

\title{Negative-index metamaterial at 780\,nm wavelength}

\author{G. Dolling and M. Wegener}
\affiliation{Institut f\"ur Angewandte Physik and DFG-Center for Functional Nanostructures (CFN), Universit\"at Karlsruhe (TH), D-76131 Karlsruhe, Germany}

\author{C.\,M. Soukoulis$^*$}
\affiliation{Ames Laboratory and Department of Physics and Astronomy, Iowa State University, Ames, Iowa 50011, U.S.A.}

\author{S. Linden}
\affiliation{Institut f\"ur Nanotechnologie, Forschungszentrum Karlsruhe in der Helmholtz-Gemeinschaft, D-76021 Karlsruhe, Germany}

\begin{abstract}

\noindent We further miniaturize a recently established silver-based negative-index metamaterial design.
By comparing transmittance, reflectance and phase-sensitive time-of-flight experiments to theory,
we infer a real part of the refractive index of -0.6 at 780\,nm  wavelength -- which is visible in the laboratory.\\
\copyright  2006 Optical Society of America
\end{abstract}

\maketitle

\newpage
\begin{figure}
\centerline{\scalebox{0.6}{\includegraphics[width=8.1cm,keepaspectratio]{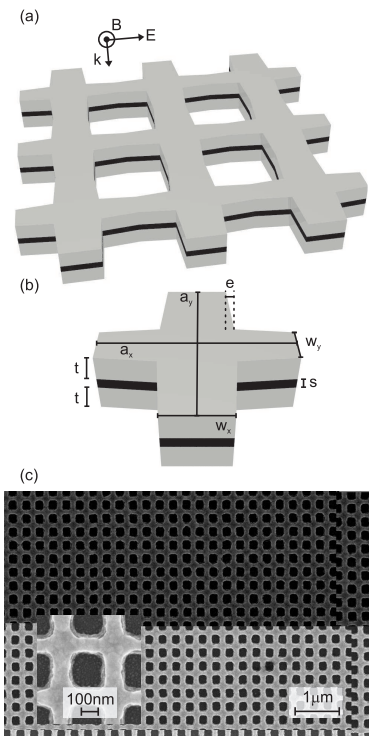}}}
\caption{ (a) Scheme of the metamaterial and polarization configuration. (b) Unit cell
of the structure with definition of parameters: lattice constant
$a_x=a_y=300\,$nm, $w_x=102\,$nm, $w_y=68\,$nm, $t=40\,$nm, $s=17\,$nm, and $e_x=e_y=e=8\,$nm.
The latter parameter describes small deviations from rectangular shape.
(c) Top-view electron micrograph of the sample employed in Figs.\,2 and 3. The inset shows a magnified view.}
\end{figure}
Photonic metamaterials are tailored artificial optical materials composed of sub-wavelength metallic
building blocks that can be viewed as nano-scale electronic circuits. These building blocks or ``photonic atoms''
are densely packed into an effective material such that the operation wavelength $\lambda$ is ideally much
larger than the lattice constant $a$. Along these lines, highly unusual material properties become accessible, e.g.,
a negative index of refraction \cite{2,4}, which has recently reached operation wavelengths of $2\,\rm \mu m$ \cite{10},
$1.5\,\rm \mu m$ \cite{11}, $1.5\,\rm \mu m$ \cite{12}, and $1.4\,\rm \mu m$ \cite{13}.
In this letter, we demonstrate a negative index of refraction at the red end of the visible spectrum (780\,nm wavelength) for the first time.

The physics of the particular sample/circuit design \cite{16} used and miniaturized here has been described
previously in work at lower frequencies.\cite{10,12,13} In brief, for the polarization configuration
shown in Fig.\,1 (a), the metamaterial can be viewed as composed of two sets of sub-circuits or ``atoms'':
(i) A coil with inductance $L$ in series with two capacitors with net capacitance $C$ as an $LC$ circuit,
providing a magnetic resonance at the $LC$ resonance frequency.\cite{14} (ii) Long metallic wires, acting like
a diluted metal below the effective plasma frequency of the arrangement.\cite{23}
The negative magnetic permeability from (i) and the negative electric permittivity from (ii) lead to a negative index of refraction \cite{2,4}.
We use silver as constituent
material because silver is known to introduce significantly lower losses\cite{24} than, e.g., gold at visible
frequencies. The choice of the dielectric spacer material is uncritical, we use $\rm MgF_2$.
The design parameters have carefully been optimized regarding optical performance on the computer.
Results from the best fabricated sample are shown here.

Fabrication employs standard electron-beam lithography, electron-beam evaporation of the constituent materials, and a
lift-off procedure. All samples are located on glass substrate, coated with a 5-nm thin film of indium-tin-oxide
(ITO) to avoid charging effects in the electron-beam-writing process (the ITO layer is irrelevant for the optical performance).
The electron micrograph of the best sample
($100\,\rm  \mu m \times 100\,\rm  \mu m$ footprint) shown in Fig.\,1 (c) reveals good large-scale homogeneity as
well as 68-nm minimum lateral feature size at 97-nm thickness of the Ag-MgF$_2$-Ag sandwich. This
aspect ratio (i.e., height/width) exceeding unity poses significant fabrication challenges. Compared
with our previous choice of parameters at lower frequencies,\cite{12,13} especially the relative thickness
of the metal wires oriented along the electric-field vector (i.e., the ratio $w_y/a_y$ in Fig.\,1 (b)) had
to be increased. This step increases the effective plasma frequency, which needs to be above the
operation frequency in order to obtain a negative electric permittivity.
\begin{figure}[h]
\centerline{\scalebox{1.0}{\includegraphics[width=7.8cm,keepaspectratio]{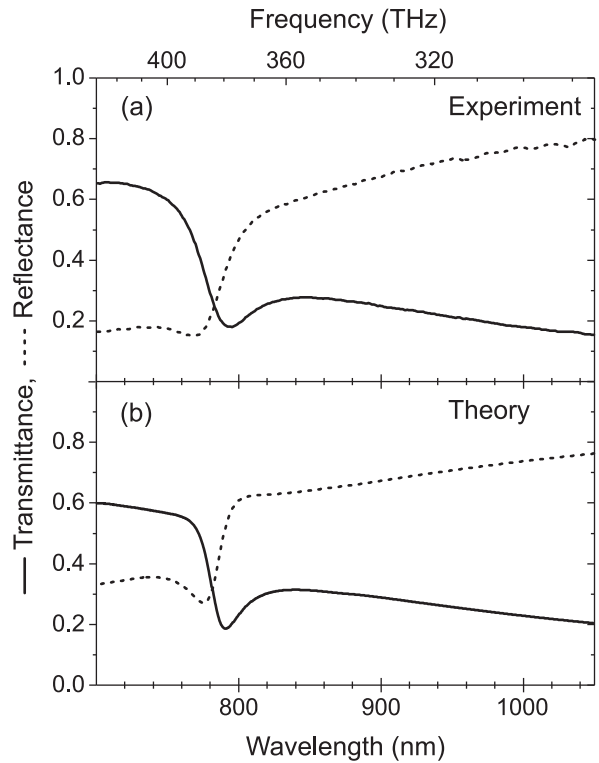}}}
\caption{ (a) Measured transmittance (solid) and reflectance (dashed) spectrum of the negative-index
metamaterial described in Fig.\,1 for the polarization configuration of Fig.\,1 (a). (b) Corresponding theoretical calculation.
The exact same parameters are used in the calculations depicted in Fig.\,3.}
\end{figure}
Figure 2 (a) shows measured normal-incidence intensity transmittance and reflectance spectra (taken with 5 degrees half-opening angle) of this
metamaterial sample. The bare glass substrate and a silver mirror, respectively, serve as reference.
Shown in Fig.\,2 (b) is the corresponding theoretical result based on numerical three-dimensional
finite-difference time-domain calculations using the commercial software package {\it CST MicroWave
Studio}. The geometrical parameters have already been indicated in Fig.\,1 (b), optical material
parameters taken are: MgF$_2$ refractive index $n_{\rm MgF_2}=1.38$, glass substrate index $n_{\rm substrate}=1.5$, and
the Drude model for silver with plasma frequency $\omega_{\rm pl}=1.37 \times 10^{16} / s$ and damping or collision frequency
$\omega_{\rm col} = 9\times 10^{13} / s$. At the frequencies of interest here, the Drude model is an adequate description of
the actual silver dielectric function.\cite{24} The quoted damping has been chosen to match the experiment
in the present work and especially comprises broadening effects due to any type of sample imperfection (e.g., granularity of the metal film or
inhomogeneous broadening).
The chosen damping is three times larger than the literature value.\cite{24} Importantly, the exact same set of
parameters will also be used below for the theoretical analysis of the interferometric experiments
as well as for the effective-parameter retrieval.
\begin{figure}[h]
\centerline{\scalebox{1.0}{\includegraphics[width=7.9cm,keepaspectratio]{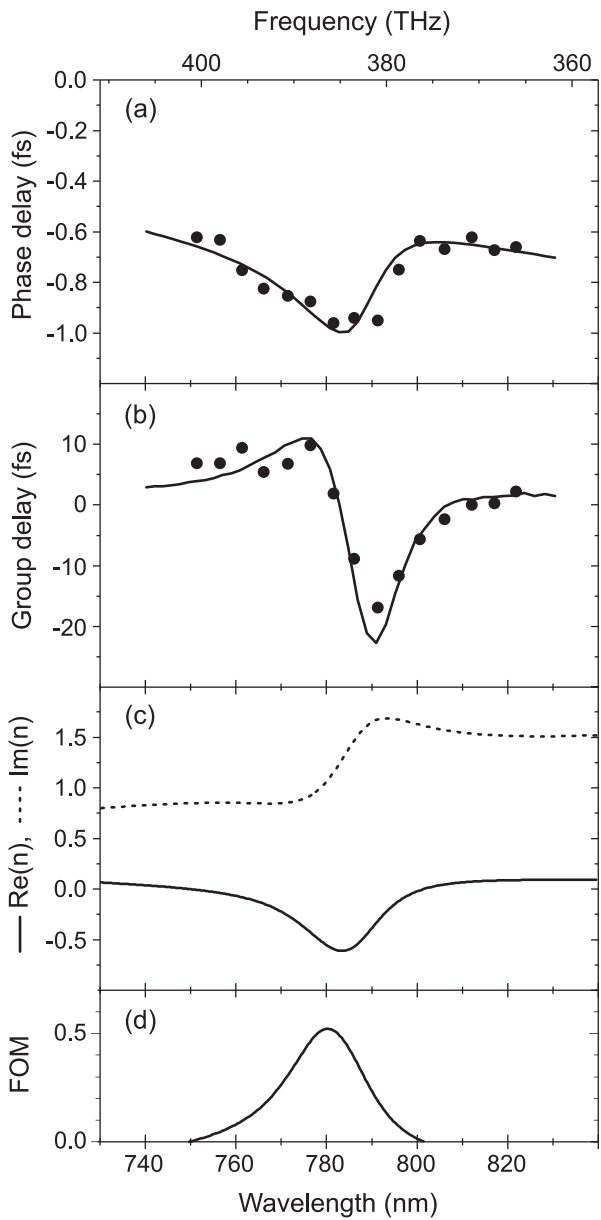}}}
\caption{ (a) Measured (dots) phase delay versus laser center wavelength for a
pulse propagating through the metamaterial sample characterized in Figs.\,1 and 2 and for
the polarization configuration depicted in Fig.\,1 (a). The solid curve is the corresponding
theoretical calculation. (b) Group delay versus wavelength. (c) Retrieved real (solid) and imaginary (dashed)
part of the effective refractive index $n$. (d) Resulting figure of merit ${\rm FOM}=-{\rm Re}(n)/{\rm Im}(n)$.
The identical set of sample parameters is used in all calculations shown in Figs.\,2 and 3.}
\end{figure}
An unambiguous determination of effective material parameters, especially of the phase velocity $c=c_0/n$,
additionally requires phase-sensitive experiments. The details and the errors of the phase-sensitive ``time-of-flight''
experiment based on a compact and passively stable Michelson interferometer have been described previously
by us\cite{12}, albeit in a different wavelength regime. In essence, we record two interferograms, one with the
metamaterial sample on its glass substrate in one of the interferometer arms, and a second interferogram
with just the glass substrate by laterally moving the metamaterial out of the optical path. All mechanical
motions are computer controlled and realized by precise and calibrated piezoelectric actuators. The
corresponding shift on the interferometer time delay axis yields the phase delay due to the metamaterial.
By tuning the center wavelength of the incident Gaussian transform-limited 125\,fs pulses, derived from a
commercial mode-locked Ti:Sa laser ({\it Spectra-Physics Tsunami}), and by repeating the described procedure
for each wavelength, we measure phase-delay spectra. Simultaneously and similarly, we infer the shift
between the two Gaussian interferogram envelopes at each wavelength, which provides us with the group-delay
spectra. In essence, the group-delay spectrum is the spectral derivative of the phase-delay spectrum.
Thus, the group-delay spectrum sensitively depends on the damping. Corresponding data (dots) are shown
in Fig.\,3 (a) and Fig.\,3 (b) together with numerical calculations (solid curves), in which we derive the
interferograms from the complex sample electric-field transmittance for the femtosecond pulse parameters as
in the experiment and then proceed with the analysis as in the experiment. Clearly,
all effects due to the finite spectral width of the pulses are appropriately accounted for in this manner.
Finally, we retrieve\cite{25} the effective material parameters from theory and depict them in Fig.\,3 (c). These ``retrieved'' parameters refer
to a fictitious homogeneous film on the glass substrate with a thickness identical to that of the metamaterial
($d=97\,$nm) and complex transmittance and reflectance properties strictly identical to those of the metamaterial
on the glass substrate. From the increasing relative importance of the imaginary part of the silver dielectric function for frequencies
(even remotely) approaching the plasma frequency, one expects increased losses.
Indeed, the figure of merit, FOM, shown in Fig.\,3 (d) and defined via ${\rm FOM}=-{\rm Re}(n)/{\rm Im}(n)$, is ${\rm FOM}=0.5$ at best, while
best values of ${\rm FOM}=3$ have recently been achieved at $1.4\,\rm \mu m$ wavelength by us.\cite{13} Still,
the obtained value of ${\rm FOM}=0.5$ is comparable to previous work at longer wavelengths.\cite{10,11,12}

Obviously, the experimental results agree well with theory, which consistently describes transmittance (solid curves in Fig.\,2),
reflectance (dashed curves in Fig.\,2), phase-delay (Fig.\,3 (a)), as well as group-delay spectra (Fig.\,3 (b)) -- all with one set of parameters. Thus, the
effective material parameters, especially the negative real part of $n$ (Fig.\,3 (c)), retrieved from the
same theory and the same parameters can be considered as very trustworthy. A determination on the basis
of, e.g., intensity reflectance spectra alone would be ambiguous and not at all reliable based on our experience.

In conclusion, we have demonstrated a metamaterial with an effective real part of the index of refraction
of $-0.6$ at around $780\,\rm nm$ wavelength. This wavelength can easily be seen with the naked eye in our laser experiments.
Our work goes beyond previous work in the visible\cite{17} which showed a
negative magnetic permeability. Phase-sensitive ``time-of-flight'' experiments give direct experimental
evidence for the negative phase velocity of light. Studying the group velocity in parallel provides a sensitive consistency check of our analysis.

We acknowledge support by the Deutsche Forschungsgemeinschaft (DFG) and the State of Baden-W\"urttemberg
through the DFG-Center for Functional Nanostructures (CFN) within subproject A1.5. The research of S. L.
is further supported through a ``Helmholtz-Hochschul-Nachwuchsgruppe'' (VH-NG-232), that of C. M. S. by the
Alexander von Humboldt senior-scientist award 2002, by Ames Laboratory (Contract No. W-7405-Eng-82),
EU projects PHOREMOST, METAMORPHOSE and DARPA (HR0011-05-C-0068).

$^*$ C.M.S. is also at Institute of Electronic Structure and Laser at FORTH and Dept. of Materials Science and Technology, Univ. of Crete,
Heraklion, Crete, Greece.


\end{document}